\documentclass[12pt,showpacs,showkeys,amsmath,aps,prc]{revtex4}
\usepackage{subfigure}
\usepackage{supertabular}
\usepackage{color,graphicx}
\usepackage[bookmarksopen]{hyperref}

\def  \o    {\omega}

\def  \s    {\sigma}

\def  \p    {\pi}

\def  \m    {\mu}
\def  \n    {\nu}

\def  \f    {\frac}
\def  \lt   {\left}
\def  \rt   {\right}
\def  \th   {\theta}

\def  \ra   {\rightarrow}

\def  \veps {\varepsilon}

\def  \gm   {\gamma^\mu}

\def  \del  {\partial}
\def  \cl   {{\cal{L}}}

\def  \mn   {\mu \nu}

\def  \bef  {\begin{figure}}
\def  \eef  {\end{figure}}
\def  \be   {\begin{equation}}
\def  \ee   {\end{equation}}
\def  \ba   {\begin{array}}
\def  \ea   {\end{array}}
\def  \bea  {\begin{eqnarray}}
\def  \eea  {\end{eqnarray}}
\def  \beq  {\begin{eqnarray}}
\def  \eeq  {\end{eqnarray}}
\def  \nn   {\nonumber}
\def  \bd   {\begin{displaymath}}
\def  \ed   {\end{displaymath}}
\def  \bse  {\begin{subequations}}
\def  \ese  {\end{subequations}}
\def  \bwt  {\begin{widetext}}
\def  \ewt  {\end{widetext}}

\def  \ba   {{\bf{a_1}}}

\begin{document}

\title{ Fermi Liquid parameters for dense nuclear matter in Effective
Chiral Model }

\author {Kausik \surname {Pal}}
\email {kausik.pal@saha.ac.in}
\author{Subhrajyoti \surname {Biswas}}
\author {Abhee K. \surname {Dutt-Mazumder}}
\affiliation {High Energy Physics Division, Saha Institute of Nuclear Physics,
 1/AF Bidhannagar, Kolkata 700064, India.}
\medskip

\begin{abstract}
We calculate relativistic Fermi liquid parameters (RFLPs) for the 
description of the properties of dense nuclear matter (DNM) using 
Effective Chiral Model. Analytical expressions of Fermi liquid parameters
(FLPs) are presented both for the direct and exchange contributions. We 
present a comparative study of perturbative calculation with mean field 
(MF) results. Moreover we go beyond the MF so as to estimate the pionic
contribution to the FLPs. Finally, we use these parameters to estimate some 
of the bulk quantities like incompressibility, sound velocity, symmetry 
energy etc. for DNM interacting via exchange of $\s$, $\o$ and $\p$ meson. 
In addition, we also calculate the energy densities and the binding energy 
curve for the nuclear matter. Results for the latter have been found to be consistent with two loop calculations reported recently within the same model.
\end{abstract}
\vspace{0.08 cm}

\pacs {21.65.-f, 13.75.Cs, 13.75.Gx, 21.30.Fe}

\keywords{Landau parameters, Thermodynamic quantities, Exchange energy.}

\maketitle

\section{introduction}

One of the most exciting field of contemporary nuclear research has 
been the studies of the properties of dense nuclear matter (DNM). 
Such studies are important both in the context of laboratory experiments 
and nuclear astrophysics. Therefore, several attempts have been made
in recent years to ascertain the properties of nuclear system at densities 
higher than the normal nuclear matter densities 
\cite{chin76,speth80,horowitz83,friman96,friman99,song01,holt07}.

The suitable description of nuclear matter at such high densities
is provided by Quantum Hadrodynamics (QHD) \cite{serot92}. 
Historically, QHD was developed by Walecka \cite{walecka74,chin74,vol16} 
to study the properties of neutron star where the nucleons are assumed to 
interact via the exchange of $\sigma$ and $\omega$ mesons. In this model
starting with interacting Lagrangian the relativistic field equations are
solved by making MF approximation where the meson fields 
are replaced by their vacuum expectation values.
Subsequently, starting from the same
model Chin developed a full diagrammatic scheme and showed
that the MF results can be obtained by making Hartree approximation
{\em i.e.} by retaining only the direct terms in a relativistic field 
theoretic approach \cite{chin77}. In the same work, it was also shown 
how exchange corrections can be made and analytical expressions can be 
found for the energy density and related quantities by making some 
long range approximation for the $n-n$ interaction. Since
then the QHD has undergone a series of developments which we do not 
discuss here but refer the reader to ref.\cite{serot79,kapusta81,
serot82,furnstahl87,furnstahl93,furnstahl95,furnstahl96}.

The most recent model which we use here for the description of dense
nuclear system is provided by the Chiral Effective Field theory (chEFT)
\cite{furnstahl89,furnstahl97}.
It might be recalled here, that, in such a framework, the explicit 
calculation of the Dirac vacuum is not required, rather, on the contrary,
here, the short distance dynamics are absorbed into the parameters of 
the theory adjusted phenomenologically by fitting empirical data.
For detail discussion refer the reader to ref.
\cite{furnstahl97,serot97,hu07,biswas08}.  
Recently this model has been applied \cite{hu07} to 
calculate the exchange corrections by evaluating nucleon loops involving
$\sigma$, $\omega$ and $\pi$ as intermediate states, which we 
address here partly. 

Our approach here is to study the dense nuclear system in terms
of relativistic Fermi liquid parameters (RFLPs). Such an extention of the
Fermi Liquid theory \cite{pines_book,baym_book} was first made by 
Baym and Chin in ref.\cite{baym76}.
It should, however, be noted that the calculation presented in
ref.\cite{baym76} were performed perturbatively where the original QHD
model was used. It should, however, be noted that the first application
of Fermi liquid theory (FLT) to study the nuclear system was due to Migdal 
\cite{migdal78} who used FLT to investigate the properties of unbound 
nuclear matter and finite nuclei \cite{mig_book}. FLT also provides 
theoretical foundation for the nuclear shell model \cite{mig_book} 
as well as nuclear dynamics of low energy excitations
\cite{baym_book,krewald88}. The connection between Landau, 
Brueckner-Bethe and Migdal theories was discussed in ref.\cite{brown71}. 
While these are all non-relativistic calculations, the relativistic 
calculations involving RFLT are rather limited.

After the original work of \cite{baym76},  the relativistic problem was
revisited in \cite{matsui81} where one starts 
from the expression of energy density in presence of scalar and vector 
meson MF and takes functional derivatives to determine the FLPs. The results 
are found to be qualitatively different than the perturbative results 
\cite{baym76,matsui81}. Moreover, 
besides $\sigma$ and $\omega$ meson, ref.\cite{matsui81} 
also includes the $\rho$ and $\pi$ meson and the model adopted was originally
proposed by Serot that incorporates pion into the Walecka model. The latter,
however, do not contribute to the parameters presented in \cite{matsui81}
as the calculation was restricted only to the MF level where pion
fails to contribute. On the other hand, the Migdal parameters 
using one-boson-exchange models of the nuclear force calculated in 
ref.\cite{celenza_book,anastasio83}, in which a comparison of relativistic
and non-relativistic results have also been studied. 

In the present work, we use a model, where we have pions and we extend
the calculation beyond MF to include the pionic contributions
into the FLPs. Furthermore, we evaluate and compare the perturbative
results with MF approximated results within the framework of the 
present model. In addition we also calculate various physical quantities
like incompressibility, sound velocity and symmetry energy etc. Moreover,
the results are compared whenever possible with the previous calculations 
by taking suitable limits. For instance, the exchange energy, we 
compare results calculated within the present scheme with a more direct
evaluation of the loop diagrams like in ref.\cite{hu07}.

This paper is organized as follows. In Sec.II, we will depict brief 
outline of the formalism of FLT. We find the analytic expressions for 
the FLPs both for direct and exchange contributions in Sec.III. 
Subsequently, we determine chemical potential, energy density and various
other thermodynamic quantities like incompressibility and sound velocity.
Sec.IV, is devoted to calculate isovector LPs to which involves the 
$\pi$ meson contribution, and used to express the symmetry energy.


\vskip 0.4in
\section {Formalism}

In FLT total energy $E$ of an interacting system is the functional of 
occupation number $n_{p}$ of the quasi-particle states of momentum $p$. 
The excitation of the system is equivalent to the change of occupation 
number by an amount $\delta n_{p}$. The corresponding energy of the system
is given by \cite{baym_book,baym76},

\beq{\label {total_energy}}
E&=&E^{0}+\sum_{s}\int\frac{d^3{p}}{(2\pi)^3}
\varepsilon_{ps}^{0}\delta n_{ps}
+\frac{1}{2}\sum_{ss'}\int\frac{d^3{p}}{(2\pi)^3}\frac{d^3{p'}}{(2\pi)^3}
f_{ps,p's'}
\delta n_{ps}\delta n_{p's'},
\eeq 

where $E^0$ is the ground state energy and $s$ is the spin index, 
and the quasi-particle energy can be written as,

\beq\label{quasi_energy}
\veps_{ps}=\veps_{ps}^{0}+\sum_{s'}\int\frac{d^3{p'}}{(2\pi)^3}f_{ps,p's'}
\delta n_{p's'},
\eeq

where $\veps_{ps}^{0}$ is the non-interacting single particle energy.
The interaction between quasi-particles is given by $f_{ps,p's'}$, which 
is defined to be the second derivative of the energy functional with respect
to occupation functions,

\beq\label{quasi_interac}
f_{ps,p's'}=\frac{\delta^{2}E}{\delta{n}_{ps}~\delta{n}_{p's'}}. 
\eeq

Since quasi-particles are well defined only near the Fermi surface, one assumes

\beq
\left.\begin{array}{lll} 
&\veps_{p}&=\mu+v_{f}(p-p_{f})\\ 
{\rm and~~~} & p&\simeq p'\simeq p_{f}.
\end{array}
\right\}
\eeq

Then LPs $f_{l}$s are defined by the Legendre expansion of $f_{ps,p's'}$ as
\cite{baym_book,baym76},

\beq\label{landau_para}
f_l=\frac{2l+1}{4}\sum_{ss'}\int\frac{d\Omega}{4\pi}P_{l}(\cos\theta)f_{ps,p's'},
\eeq

where $\theta$ is the angle between $p$ and $p'$, both taken to be on the 
Fermi surface, and the integration is over all directions of $p$ \cite{baym76}. 
We restrict ourselves for $l\le 1$ i.e. $f_{0}$ and $f_{1}$, since higher $l$ 
contribution decreases rapidly.

\begin{figure}[htb]
\begin{center}
\includegraphics[scale=0.65,angle=0]{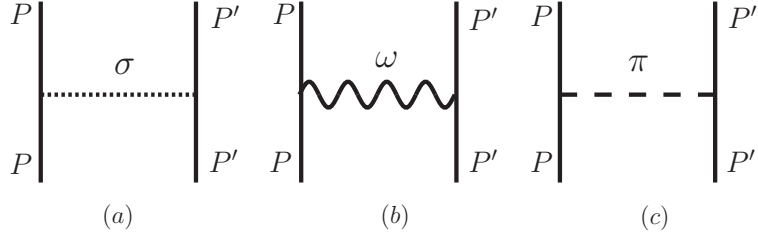}
\caption{Schematic diagrams of direct contribution to forward scattering
amplitude. Nucleons are represented by solid lines. $\sigma$, $\omega$
and $\pi$ mesons are denoted by dotted, wavy and dashed lines respectively.}
\label{dir_jax}
\end{center}
\end{figure}

\begin{figure}[htb]
\begin{center}
\includegraphics[scale=0.65,angle=0]{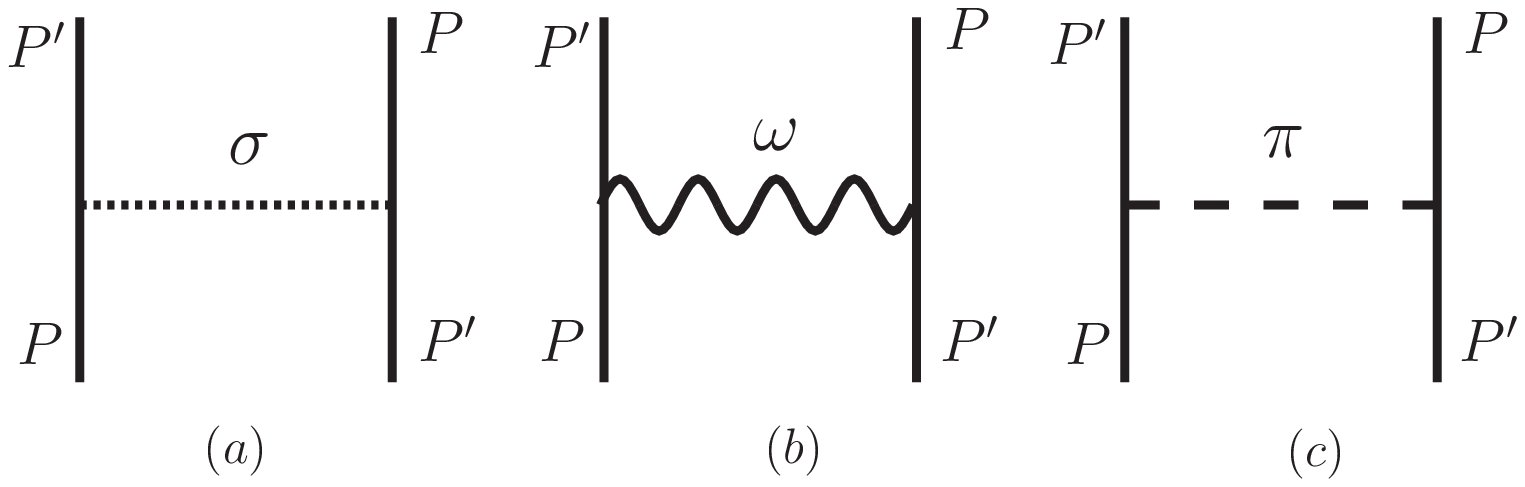}
\caption{Schematic diagrams of exchange contribution to forward scattering
amplitude. Nucleons are represented by solid lines. $\sigma$, $\omega$
and $\pi$ mesons are denoted by dotted, wavy and dashed lines respectively.}
\label{ex_jax}
\end{center}
\end{figure}

Now the Landau Fermi liquid interaction $f_{ps,p's'}$ is related to the
two particle forward scattering amplitude via \cite{baym_book,baym76},

\beq
f_{ps,p's'}&=&\frac{M}{\veps_{p}^0}\frac{M}{\veps_{p'}^0}
{\cal M}_{ps,p's'},
\eeq

where $M$ is the mass of the nucleon and the Lorentz invariant matrix  
${\cal M}_{ps,p's'}$ consists of the usual direct and exchange amplitude, 
which may be evaluated directly from the relevant diagrams as shown in 
Fig.~\ref{dir_jax} and Fig.~\ref{ex_jax}.

\vskip 0.2in

\vskip 0.2in

The spin averaged scattering amplitude  $(f_{pp'})$ is given by \cite{baym76},

\beq\label{fermi_inter}
f_{pp'}&=&\frac{1}{4}\sum_{ss'}\frac{M}{\veps_{p}^0}
\frac{M}{\veps_{p'}^0}{\cal M}_{ps,p's'}.
\eeq 

The dimensionless LPs are 
$F_{l}=N(0)f_{l}$, where $N(0)$ is the density of states at the Fermi surface 
defined as,

\beq\label{dens_of_state} 
N(0)&=&\sum_{s}\int\frac{\rm d^3{p}}{(2\pi)^3}
\delta(\veps_{ps}-\mu)\nn\\
&=&\frac{g_{s}g_{I}p_{f}^2}{2\pi^2}\left(\frac{\del p}{\del\veps_{p}}
\right)_{p=p_{f}}\nn\\
&\simeq&\frac{g_{s}g_{I}p_{f}\veps_{f}}{2\pi^2}.
\eeq

Here $g_{s},g_{I}$ are 
the spin and isospin degeneracy factor respectively.

In the above expression $(\del p/\del\veps_{p})_{p=p_{f}}$
is the inverse Fermi velocity $(v_{f}^{-1})$ related to the FL parameter $F_{1}$,
\beq\label{inv_vel}
v_{f}^{-1}=(\del p/\del\veps_{p})_{p=p_{f}}=(\mu/p_{f})(1+F_{1}/3).
\eeq

With the help of Eq.(\ref{dens_of_state}) and Eq.(\ref{inv_vel})
one writes {\cite{matsui81}}
 
\beq\label{relative}
\veps_{f}=\mu(1+\frac{1}{3}F_{1}).
\eeq

To compare Eq.(\ref{inv_vel}) and Eq.(\ref{relative}) with the well known 
non-relativistic expressions one has to put $\veps_{f}=M^*$ and $\mu=M$.


\vskip 0.4in
\section {Chiral Lagrangian and Landau parameters}

We adopt the non-linear chiral model to calculate the FLPs and 
consequently estimate various quantities of physical interest like 
effective chemical potential, sound velocity, incompressibility, 
symmetry energy etc. Here all the fields are treated relativistically 
\cite{baym76}. By retaining only the 
lowest order terms in the pion fields, one obtains the following Lagrangian
from the chirally invariant Lagrangian \cite{furnstahl97,hu07}:

\beq
\cl &=& \bar{\Psi}\left[\gm(i\del_{\mu}-g_{\o}\o_{\mu})-i\frac{g_{A}}{f_{\pi}}
\gm\gamma_{5}\del_{\mu}\underline{\pi}-(M-g_{\s}\Phi_{\s})\right]\Psi\nn\\&&
+\frac{1}{2}\del^{\mu}\Phi_{\s}\del_{\mu}\phi_{\s}-\frac{1}{2}m_{\s}^2\Phi_{\s}^2
-\frac{1}{4}\o^{\mn}\o_{\mn}+\frac{1}{2}m_{\o}^2\o^{\m}\o_{\m}
+\frac{1}{2}\del^{\m}{\vec\Phi_{\p}}\cdot\del_{\m}{\vec\Phi_{\p}}
-\frac{1}{2}m_{\p}^2{\vec\Phi_{\p}^2}\nn\\&&
+\cl_{NL}+\delta\cl,
\eeq

where $\o_{\mn}=\del_{\m}\o_{\n}-\del_{\n}\o_{\m}$,
$\underline{\p}=\frac{1}{2}({\vec\tau}\cdot{\vec\Phi_{\p}})$ and 
${\vec\tau}$ is the isospin index. Here $\Psi$ is the nucleon
field and $\Phi_{\s}$, $\o_{\m}$ and ${\vec\Phi_{\p}}$ are the meson fields 
(isoscalar-scalar, isoscalar-vector and isovector-pseudoscalar respectively).
The terms $\delta\cl$ and $\cl_{NL}$ contain the non-linear and counterterms respectively (for explicit expression see{\cite{hu07}}). Note that in this work,
the convention of {\cite {kapusta81}} is used.

\subsection{Perturbative calculation}

Let us calculate the LPs perturbatively due to the exchange of scalar 
and vector mesons between the nucleons \cite{baym76}. The direct 
contribution ({\em see } Fig.~\ref{dir_jax}) is given by \cite{baym76}

\beq\label{dir_int}
\left.\begin{array}{ll}
f_{pp'}^{dir,\s}&=-\frac{g_{\sigma}^2}{m_{\sigma}^2}\frac{M^2}
{\veps_{p}^0\veps_{p'}^0}\\
f_{pp'}^{dir,\o}&=\frac{g_{\omega}^2}{m_{\omega}^2}\frac{P.P'}
{\veps_{p}^0\veps_{p'}^0},
\end{array}
\right\}
\eeq

where  $\veps_{p}^0=\sqrt{p^2+M^2}$. Now with the help of Eq.(\ref{landau_para}) and Eq.(\ref{dir_int}), the LPs become 

\beq\label{f0_dir}
\left.\begin{array}{ll}
f_{0}^{dir,\s}&=-\f{g_{\s}^2}{m_{\s}^2}\f{M^{2}}{\veps_{f}^2}\\
f_{0}^{dir,\o}&=\f{g_{\o}^2}{m_{\o}^2},
\end{array}
\right\}
\eeq

and
 
\beq\label{f1_dir}
\left.\begin{array}{ll}
f_{1}^{dir,\s}&=0\\
f_{1}^{dir,\o}&=-\f{g_{\o}^2}{m_{\o}^2}\frac{p_{f}^2}{\veps_{f}^2}.
\end{array}
\right\}
\eeq

One may neglect the contribution of $f_{1}^{dir,\o}$ as discussed in
ref.\cite{speth80}. A better approach was developed by Matsui \cite{matsui81}
where the magnetic interaction is included which reduces the value of
$f_{1}^{dir,\o}$.

One may now, for the direct contribution plug in $f_{pp'}^{dir,\s}$ and 
$f_{pp'}^{dir,\o}$ in Eq.(\ref{total_energy}) and Eq.(\ref{quasi_energy}) 
to obtain the energy density and the SPE spectrum, respectively. The SPE 
spectrum is given by \cite{baym76}

\beq\label{dir_single}
\veps_{p}^{dir}&=&\veps_{p}^{0}+\frac{g_{\o}^2}{m_{\o}^2}\rho
-\frac{g_{\s}^2}{m_{\s}^2}\frac{M}{\veps_{p}^0}n_{s}.
\eeq

Here $\rho $ and $n_{s}$ are the baryon and scalar density given by

\beq\label{baryon_den}
\rho&=&g_{s}g_{I}\frac{p_{f}^3}{6\p^2},
\eeq

and

\beq\label{scalar_den}
n_{s}&=&g_{s}g_{I}\frac{M}{4\p^2}
\left[p_{f}\veps_{f}-M^2\ln\left(\frac{p_{f}+\veps_{f}}{M}\right)\right].
\eeq

The energy density for direct contribution is \cite{baym76}

\beq\label{dir_totalE}
E^{dir}&=&E^0+\frac{1}{2}\frac{g_{\o}^2}{m_{\o}^2}\rho^2
-\frac{1}{2}\frac{g_{\s}^2}{m_{\s}^2}n_{s}^2.
\eeq

The chemical potential is 

\beq\label{mu_dir1}
\mu^{dir}&=&\frac{\del E^{dir}}{\del \rho}\nn\\
&=&\veps_{f}+\frac{g_{\o}^2}{m_{\o}^2}\rho-\frac{g_{\s}^2}{m_{\s}^2}n_{s}
\frac{\del n_{s}}{\del \rho}\nn\\
&=&\veps_{f}+\frac{g_{\o}^2}{m_{\o}^2}\rho
-\frac{g_{\s}^2}{m_{\s}^2}\frac{M}{\veps_{f}}n_{s}.
\eeq

One can derive the same result directly from
Eq.(\ref{dir_single}) as $\mu = \veps_p{\Big |_{p=p_f}}$. 

\subsection{FLPs in mean field model}

It is well known that in the MF approximation, one replaces the mesonic
fields by their vacuum expectation values viz.
$\sigma \ra <\sigma> = \sigma_0$, $\omega \ra <\omega> = 
\delta_{\mu 0} \omega^\mu$. The pion, however, fails to contribute at the 
MF level as  $ <\pi> = 0$.  In the MF approximation the energy density 
can be written as \cite{vol16}

\beq\label{mft_totalE}
E^{MFT}&=&\frac{1}{2}\frac{g_{\o}^2}{m_{\o}^2}\rho^2
+\frac{1}{2}\frac{g_{\s}^2}{m_{\s}^2}n_{s}^2+\sum_{i}n_{i}\sqrt{p_{i}^2+M^{*2}}.
\eeq

In the above equation $M^*$ denotes the effective nucleon mass to be 
determined self consistently \cite{matsui81,hu07}. With the help of 
Eq.(\ref{quasi_interac}), the interaction parameter takes the following 
form \cite{matsui81}

\beq\label{mft_int}
f_{pp'}^{MFT}&=&\frac{g_{\o}^2}{m_{\o}^2}-\frac{g_{\s}^2}{m_{\s}^2}\frac{M^{*2}}
{\veps_{p}^0\veps_{p'}^0}\left[1+\zeta(M^*)\right]^{-1}.
\eeq

Here $\veps_{p}^0=\sqrt{p^2+M^{*2}}$ and  

\beq\label{brk_trm}
\zeta(M^*)&=&\frac{g_{\s}^2}{m_{\s}^2}{\sum_{i}}n_{i}
\frac{p_{i}^2}{(p_{i}^2+M^{*2})^{3/2}}\nn\\
&=&\frac{g_{\s}^2}{m_{\s}^2}M^{*2}\frac{g_{s}g_{I}v_{f}}{2\p^2}
\left(1+\frac{1}{2(1-v_{f}^2)}-\frac{3}{4v_{f}}
\ln\left\vert\frac{1+v_{f}}{1-v_{f}}\right\vert\right),
\eeq

where $v_{f}={p_{f}}/{(p_{f}^2+M^{*2})^{1/2}}$, is the relativistic 
Fermi velocity. The inverse part of Eq.(\ref{mft_int})
reduces the magnitude of interaction parameter compared to what is obtained
in absence of the MF Eq.(\ref{dir_int}) \cite{matsui81,anastasio83}

The LPs as defined in ref.\cite{matsui81} are

\beq\label{f0_mft}
\left.\begin{array}{ll}
f_{0}^{MFT,\s}&=-\f{g_{\s}^2}{m_{\s}^2}\f{M^{*2}}{\veps_{f}^2}
\left[1+\frac{g_{\s}^2}{m_{\s}^2}
{\sum_{i}}n_{i}\frac{p_{i}^2}{(p_{i}^2+M^{*2})^{3/2}}\right]^{-1}\\
f_{0}^{MFT,\o}&=\f{g_{\o}^2}{m_{\o}^2}.
\end{array}
\right\}
\eeq

When we evaluated the above Eq.(\ref{f0_mft}), we neglect the 
``magnetic interaction'' between the quasiparticles which is induced 
by the microscopic currents. In presence of current density \cite{matsui81}

\beq\label{f1_mft}
f_{1}^{MFT,\o}&=&-\f{g_{\o}^2}{m_{\o}^2}\f{p_{f}^2}{\veps_{f}^2}
\left[1+\frac{g_{\o}^2}{m_{\o}^2}
{\sum_{i}}n_{i}\frac{\f{2}{3}p_{i}^2+M^{*2}}{(p_{i}^2+M^{*2})^{3/2}}\right]^{-1}.
\eeq

Clearly, the current contribution reduces the value of $f_{1}^{MFT,\o}$.
Previously we showed in Eq.(\ref{dir_single}) and Eq.(\ref{dir_totalE}) 
the SPE spectrum and energy density in absence of MF, but in presence of 
MF, SPE is given by \cite{matsui81,chin76,vol16,chin74}

\beq\label{mft_single}
\veps_{p}^{MFT}&=&\frac{g_{\o}^2}{m_{\o}^2}\rho+\sqrt{p^2+M^{*2}}.
\eeq

Therefore,

\beq\label{mu_mft}
\mu^{MFT}&=&\frac{g_{\o}^2}{m_{\o}^2}\rho+\sqrt{p_{f}^2+M^{*2}}.
\eeq

In the low density limit, Eq.(\ref{mft_single}) reduces to 
Eq.(\ref{dir_single}) as $M^*= M-\frac{g_{\s}^2}{m_{\s}^2}
{\sum_{i}}n_{i}\frac{M}{(p_{i}^2+M^{2})^{1/2}}$. 
It is to be noted that in the MF approximation scalar meson 
contribution is absorbed in the effective mass does not appear
explicitly as in Eq.(\ref{dir_single}). Another interesting 
difference is also noticed in the expressions for the total 
energy densities given by Eqs.(\ref{dir_totalE}) and 
(\ref{mft_totalE}). Note that, our MF result is consistent with 
ref.\cite{matsui81} but differs with that of ref.\cite{chin74}.


\begin{figure}[htb]
\vskip 0.2in
\begin{center}
\resizebox{8cm}{6.0cm}{\includegraphics[]{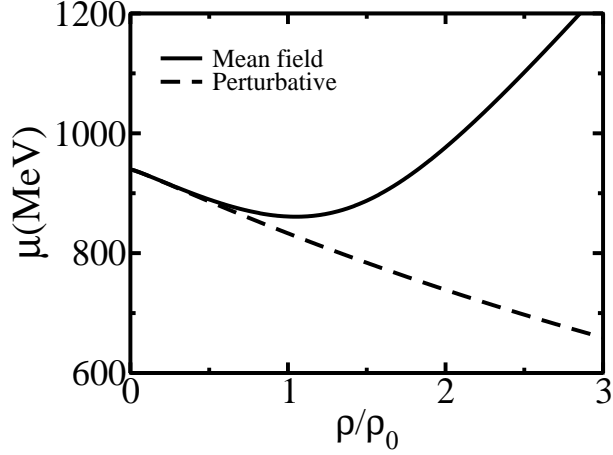}}
\caption{Chemical potential for direct contribution with $\sigma$ 
and $\omega$ meson exchange in symmetric nuclear matter. The dashed and 
solid curve represent the perturbative and MF results, respectively.}
\label{fig1}
\end{center}
\end{figure}



\begin{figure}[htb]
\begin{center}
\resizebox{8cm}{6.0cm}{\includegraphics[]{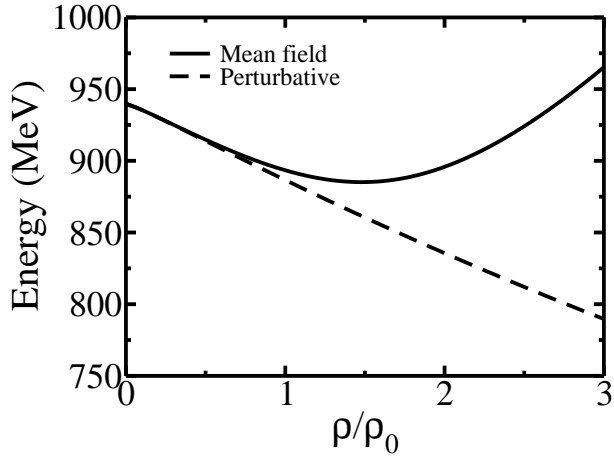}}
\caption{ Total energy from direct contribution with $\sigma$ 
and $\omega$ meson exchange in symmetric nuclear matter. The dashed and 
solid curve represent the perturbative and MF results, respectively.}
\label{fig1a}
\end{center}
\end{figure}


In Fig(\ref{fig1}) we present the comparative study of the chemical 
potential obtained perturbatively and with MF calculation. At low density 
they tend to merge, while at higher density MF results differ significantly
from the perturbative result. Numerically $\mu^{MFT}$ and $\mu^{per}$ are
given by 861.07 MeV and 832.64 MeV respectively at normal matter density 
($\rho_{0}=0.148{\rm fm^{-3}}$). For the numerical estimate we adopt the 
coupling parameter set as designated by ${\bf M0A}$ in ref.\cite{hu07}. 
In Fig(\ref{fig1a}) we compare the results for total energy obtained from perturbative and MF calculation. This also shows at low density they 
tend to merge, while at higher density MF results become larger than the  perturbative results. This is easily understood from 
Eq.(\ref{dir_totalE}) and (\ref{mft_totalE}).
At saturation density numerical values are given by 886.43 MeV and 
893.31 MeV for perturbative and MF calculation respectively.

Now we consider the exchange modification over the MF. Evaluating the exchange 
diagrams (Fig.~\ref{ex_jax}), we obtain the interaction parameter as 
\cite{baym76}

\beq
\left.\begin{array}{ll}
f_{pp'}^{ex,\s}&=-\frac{g_{\sigma}^2}{4\veps_{p}^0\veps_{p'}^0}\frac{P.P'+M^{*2}}
{(P-P')^2-m_{\sigma}^2}\\
f_{pp'}^{ex,\o}&=-\frac{g_{\omega}^2}{2\veps_{p}^0\veps_{p'}^0}\frac{P.P'-2M^{*2}}
{(P-P')^2-m_{\omega}^2}.
\end{array}
\right\}
\eeq

With the help of Eq.(\ref{landau_para}), the LPs for scalar meson 
exchange reads as

\beq\label{f0sig}
f_{0}^{ex,\sigma}&=&\frac{g_{\s}^2}{8\veps_{f}^2}\int_{-1}^{1}
\frac{p_{f}^2(1-\cos\theta)+2M^{*2}}{2p_{f}^2(1-\cos\theta)+m_{\s}^2}
{\rm d(\cos\theta)}\nn\\
&=&\frac{g_{\sigma}^2}{8\veps_{f}^2}\left[1-\left(
\frac{m_{\sigma}^2-4M^{*2}}{4p_{f}^2}\right)\ln\left(1+\frac{4p_{f}^2}{m_{\sigma}^2}
\right)\right],
\eeq

and

\beq\label{f1sig}
\frac{1}{3}f_{1}^{ex,\sigma}&=&\frac{g_{\s}^2}{8\veps_{f}^2}\int_{-1}^{1}
\frac{p_{f}^2(1-\cos\theta)+2M^{*2}}{2p_{f}^2(1-\cos\theta)+m_{\s}^2}(\cos\theta)
{\rm d(\cos\theta)}\nn\\
&=&\frac{g_{\sigma}^2}{8\veps_{f}^2}\left[\left(\frac
{m_{\sigma}^2-4M^{*2}}{2p_{f}^2}\right)\left\{1-\left(\frac{m_{\sigma}^2+2p_{f}^2}
{4p_{f}^2}\right)\ln\left(1+\frac{4p_{f}^2}{m_{\sigma}^2}\right)\right\}\right].
\eeq

Using Eqs.(\ref{f0sig}) and (\ref{f1sig}) we have

\beq\label{f0f1sig}
f_{0}^{ex,\sigma}-\frac{1}{3}f_{1}^{ex,\sigma}&=&
\frac{g_{\s}^2}{8\veps_{f}^2}\int_{-1}^{1}
\frac{p_{f}^2(1-\cos\theta)+2M^{*2}}{2p_{f}^2(1-\cos\theta)+m_{\s}^2}
(1-\cos\theta){\rm d(\cos\theta)}\nn\\
&=&\frac{g_{\sigma}^2}{8\veps_{f}^2}
\left[1-\left(\frac{m_{\sigma}^2-4M^{*2}}{2p_{f}^2}\right)\left\{1-
\frac{m_{\sigma}^2}{4p_{f}^2}\ln\left(1+\frac{4p_{f}^2}{m_{\sigma}^2}\right)
\right\}\right].
\eeq

It is this combination {\em i.e.} $f_{0}-\frac{1}{3}f_{1}$, which appears
in the calculation of chemical potential and other relevant quantities.
For massless scalar meson interaction, the above Eq.(\ref{f0f1sig})
turns out to be finite,

\beq\label{f0f1sig1}
\left(f_{0}^{ex,\sigma}-\frac{1}{3}f_{1}^{ex,\sigma}\right)_{m_{\s}\ra 0}
&=&\frac{g_{\s}^2}{8p_{f}^2}\left(1+\frac{M^{*2}}{\veps_{f}^2}\right).
\eeq

Note that in the limit $m_{\s}\ra 0$ both $f_{0}$ and $f_{1}$  are individually
diverge because of the presence of $(1-\cos\theta)$ term in the denominator.
In the massless limit such divergences are also contained in Eq.(\ref{f0sig}) 
and Eq.(\ref{f1sig}).

The dimensionless LPs $F_{0}$ and $F_{1}$ are defined as
$F_{0}=N(0)f_{0}$ and $F_{1}=N(0)f_{1}$, where $N(0)$ is the density of 
states at the Fermi surface defined in Eq.(\ref{dens_of_state}).

\beq
F_{0}^{ex,\s}&=&g_{s}g_{I}\frac{g_{\sigma}^2p_{f}}{16\pi^2\veps_{f}}\left[1-\left(
\frac{m_{\sigma}^2-4M^{*2}}{4p_{f}^2}\right)\ln\left(1+\frac{4p_{f}^2}{m_{\sigma}^2}
\right)\right],
\eeq

and

\beq
\f{1}{3}F_{1}^{ex,\s}&=&g_{s}g_{I}\frac{g_{\sigma}^2p_{f}}{16\pi^2\veps_{f}}
\left[\left(\frac
{m_{\sigma}^2-4M^{*2}}{2p_{f}^2}\right)\left\{1-\left(\frac{m_{\sigma}^2+2p_{f}^2}
{4p_{f}^2}\right)\ln\left(1+\frac{4p_{f}^2}{m_{\sigma}^2}\right)\right\}\right].
\eeq

Similarly, for vector meson exchange we have

\beq\label{f0omega}
f_{0}^{ex,\omega}&=&\frac{g_{\omega}^2}{4\veps_{f}^2}\int_{-1}^{1}
\frac{p_{f}^2(1-\cos\theta)-M^{*2}}{2p_{f}^2(1-\cos\theta)+m_{\o}^2}
{\rm d(\cos\theta)}\nn\\
&=&-\frac{g_{\omega}^2}{8\veps_{f}^2}\left[-2+
\frac{(m_{\omega}^2+2M^{*2})}{2p_{f}^2}\ln\left(1+\frac{4p_{f}^2}{m_{\omega}^2}
\right)\right],
\eeq

and

\beq\label{f1omega}
\frac{1}{3}f_{1}^{ex,\omega}&=&\frac{g_{\omega}^2}{4\veps_{f}^2}\int_{-1}^{1}
\frac{p_{f}^2(1-\cos\theta)-M^{*2}}{2p_{f}^2(1-\cos\theta)+m_{\o}^2}
(\cos\theta){\rm d(\cos\theta)}\nn\\
&=&-\frac{g_{\omega}^2(m_{\omega}^2+2M^{*2})}
{16\veps_{f}^2p_{f}^2}\left[-2+\left(1+\frac{m_{\omega}^2}{2p_{f}^2}\right)
\ln\left(1+\frac{4p_{f}^2}{m_{\omega}^2}\right)\right].
\eeq

Using Eq.(\ref{f0omega}) and Eq.(\ref{f1omega}) we have

\beq\label{f0f1ome}
f_{0}^{ex,\omega}-\frac{1}{3}f_{1}^{ex,\omega}&=&
\frac{g_{\omega}^2}{4\veps_{f}^2}\int_{-1}^{1}
\frac{p_{f}^2(1-\cos\theta)-M^{*2}}{2p_{f}^2(1-\cos\theta)+m_{\o}^2}
(1-\cos\theta){\rm d(\cos\theta)}\nn\\
&=&\frac{g_{\omega}^2}{4\veps_{f}^2}
\left[1+\frac{(m_{\omega}^2+2M^{*2})}{4p_{f}^2}\left\{-2+\frac{m_{\omega}^2}
{2p_{f}^2}\ln\left(1+\frac{4p_{f}^2}{m_{\omega}^2}\right)\right\}\right].
\eeq

In the limit $m_{\o}\ra 0$ the above Eq.(\ref{f0f1ome}) turns into

\beq\label{f0f1ome1}
\left(f_{0}^{ex,\o}-\frac{1}{3}f_{1}^{ex,\o}\right)_{m_{\o}\ra 0}
&=&\frac{g_{\o}^2}{4p_{f}^2}\left(1-\frac{2M^{*2}}{\veps_{f}}\right).
\eeq

This expression agrees with the previous calculation by Baym and Chin
\cite{baym76} who arrived at this result by direct evaluation of the 
integral by putting $m_{\o}=0$ in Eq.(\ref{f0f1ome}). Here also 
to be noted, in the limit $m_{\o}\ra 0$ both $f_{0}$ and $f_{1}$ are  
individually divergent, but the combination $f_{0}-\frac{1}{3}f_{1}$ 
is finite as observed in the case of scalar ($\s$) meson exchange.

The dimensionless LPs for vector meson exchange reads:

\beq
F_{0}^{ex,\o}&=&-g_{s}g_{I}\frac{g_{\omega}^2p_{f}}{16\pi^2\veps_{f}}\left[-2+
\frac{(m_{\omega}^2+2M^{*2})}{2p_{f}^2}\ln\left(1+\frac{4p_{f}^2}{m_{\omega}^2}
\right)\right],
\eeq

and

\beq
\f{1}{3}F_{1}^{ex,\o}&=&-g_{s}g_{I}\frac{g_{\omega}^2(m_{\omega}^2+2M^{*2})}
{32\pi^2p_{f}\veps_{f}}\left[-2+\left(1+\frac{m_{\omega}^2}{2p_{f}^2}\right)
\ln\left(1+\frac{4p_{f}^2}{m_{\omega}^2}\right)\right].
\eeq

\vskip 0.2in
\begin{figure}[htb]
\begin{center}
\resizebox{8cm}{6.0cm}{\includegraphics[]{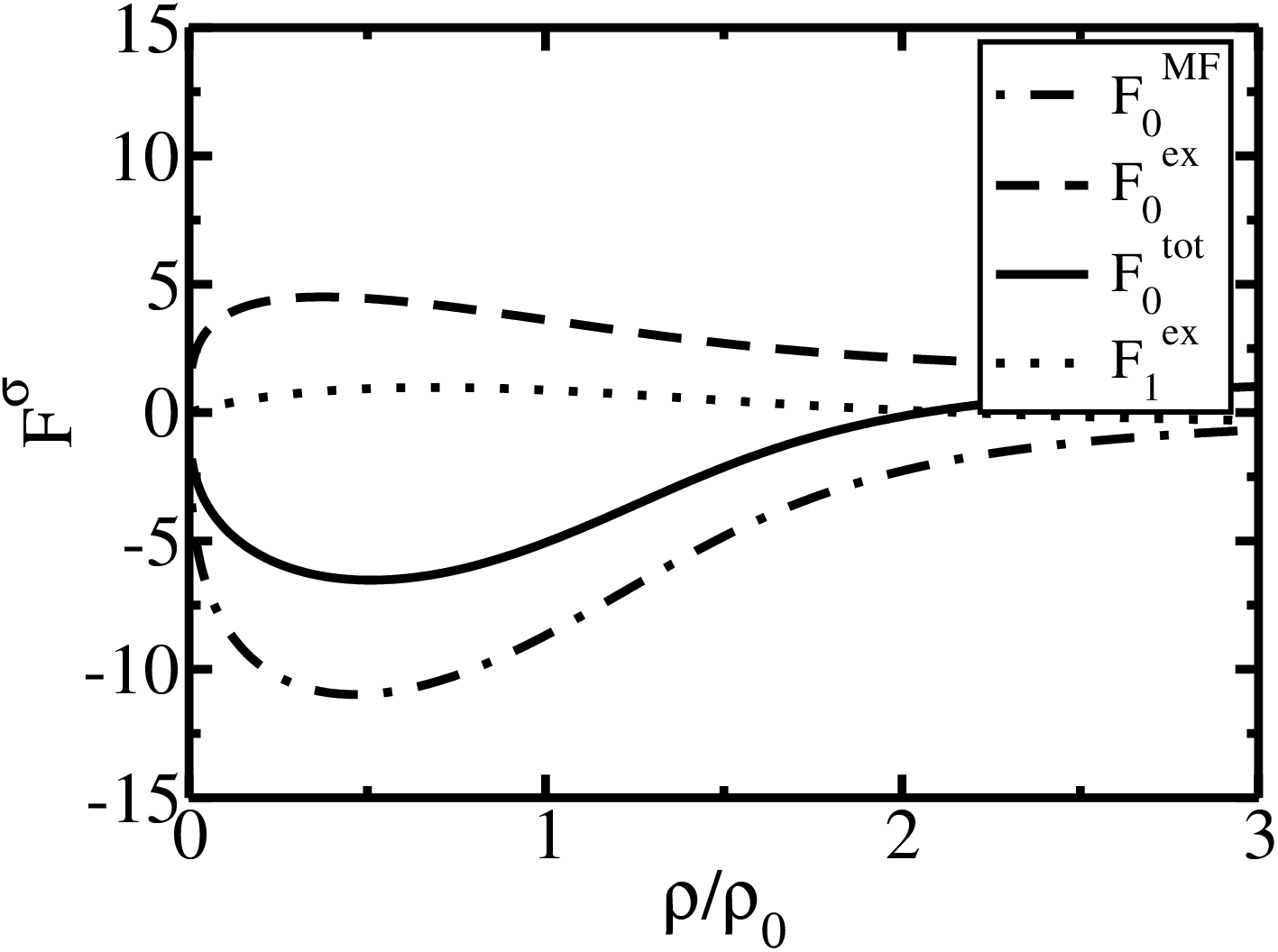}}
\caption{Dimensionless LPs in symmetric nuclear 
matter for $\sigma$ meson exchange in relativistic theory. $F_{0}^{MF}$,
 $F_{0}^{ex}$, $F_{0}^{tot}$ and $F_{1}^{ex}$ are denoted by dot-dashed, dashed, 
solid and dotted line respectively.}
\label{fig2}
\end{center}
\end{figure}



\begin{table}
\caption{Parameter sets used in this work.}
\label{table-1}
\begin{tabular}{ccccc} \hline \hline
 Meson    &~~~~    & Mass               &~~~~ & Coupling \\  \hline
  $\s$    &~~~~    & $m_{\s}/M$=0.54     &~~~~ & $g_{\s}/{4\p}$=0.7936\\
  $\o$    &~~~~    & $m_{\o}/M$=0.8328   &~~~~ & $g_{\o}/{4\p}$=0.9681\\ \hline
\hline
\end{tabular}
\end{table}


\vskip 0.2in
\begin{figure}[htb]
\begin{center}
\resizebox{8cm}{6.0cm}{\includegraphics[]{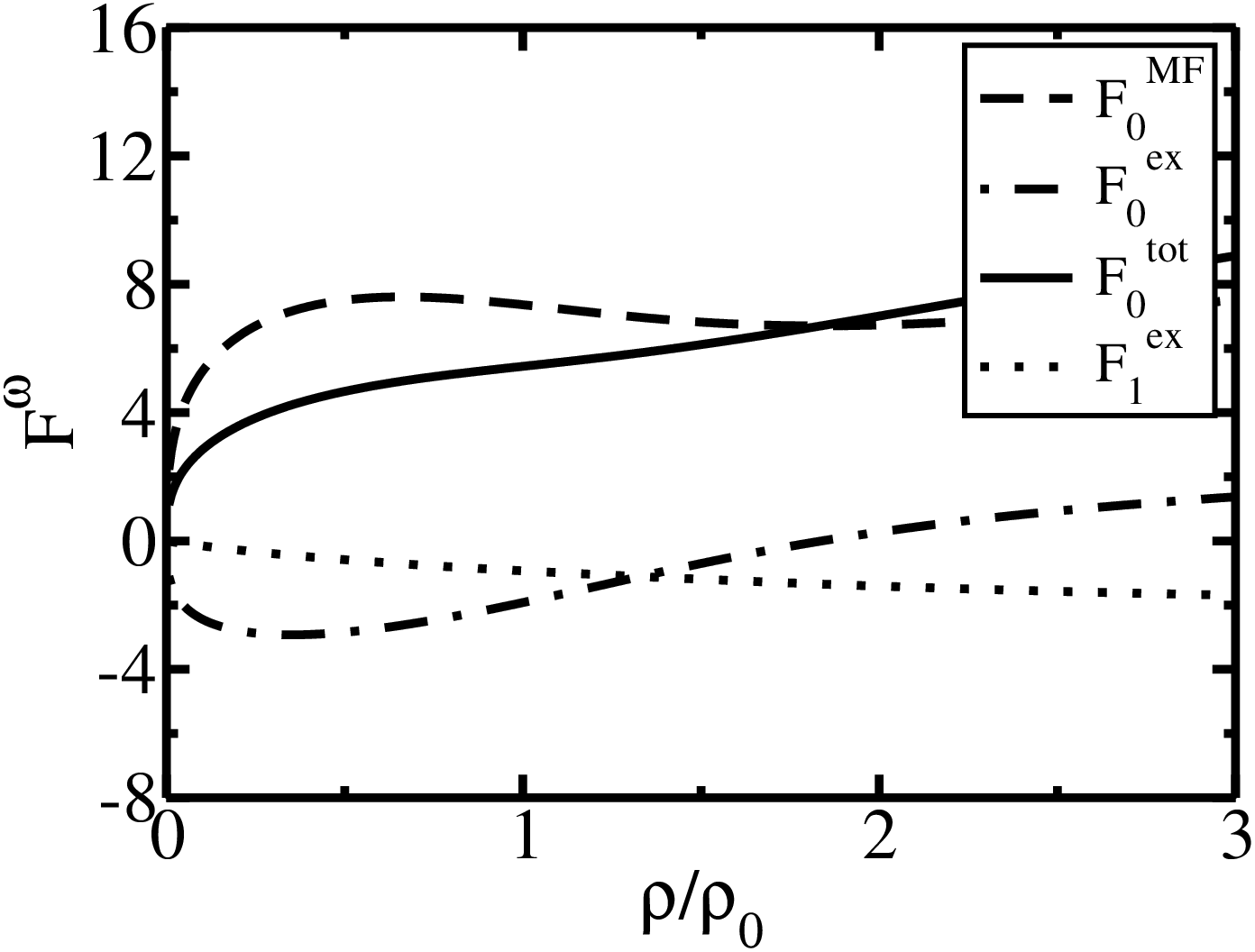}}
\caption{Dimensionless LPs for symmetric nuclear 
matter for $\omega$ meson exchange in relativistic theory. 
$F_{0}^{MF}$, $F_{0}^{ex}$, $F_{0}^{tot}$ and $F_{1}^{ex}$ are denoted by dashed, 
dot-dashed, solid and dotted line respectively.}
\label{fig3}
\end{center}
\end{figure}


\begin{table}
\caption{Dimensionless LPs for $\s$ and $\o$ exchange at 
$\rho=\rho_{0}$.}
\label{table-2}
\begin{tabular}{ccccccccc}\hline\hline
 Meson &~~~~&$F_{0}^{MF}$&~~~~&$F_{0}^{ex}$&~~~~&$F_{0}^{tot}$&~~~~&$F_{1}^{ex}$
\\ \hline
  $\s$  &~~~~& -8.65 &~~~~ &  3.61 &~~~~ &  -5.04 &~~~~  &  0.875   \\
  $\o$  &~~~~&  7.35 &~~~~ & -1.91 &~~~~ &   5.44 &~~~~  &  -0.932   \\ \hline
\hline
\end{tabular}
\end{table}


In Fig.(\ref{fig2}) and Fig.(\ref{fig3}) we present  
$F_{0}^{MF}$, $F_{0}^{ex}$, $F_{0}^{tot}$ and $F_{1}^{ex}$ as a function 
of baryon density for symmetric nuclear matter due to 
$\sigma$ and $\omega$ meson interaction respectively.  
It is to be noted, that $F_{0}^{MF}$ and $F_{0}^{ex}$ contribute 
in opposite sign for both $\s$ and $\o$ meson exchange as it is seen 
from Table~(\ref{table-2}). 
We quote few numerical values of $F^{\s}$ and $F^{\o}$ in Table~(\ref{table-2}) 
at normal matter density ($\rho_{0}=0.148 {\rm fm^{-3}}$). It is to 
be noted that the numerical estimation of $F_0$ with MF in our case, 
differs from ref.\cite{matsui81}. This is due to different coupling 
parameters in these two models.

We now proceed to calculate chemical potential due to the exchange terms 
denoted by $\mu^{ex}$. As in ref.{\cite{baym76}} we have 

\beq\label{delmu_deln}
\frac{\del\mu}{\del \rho}&=&\frac{2\pi^2}{g_{s}g_{I}p_{f}^2} 
\left(\frac{\del\veps_{p}}{\del p}\right)_{p=p_{f}}+f_{0} 
 \nn \\
&=&\frac{2\pi^2}{\mu g_{s}g_{I}p_{f}}-\frac{1}{3}f_{1}+f_{0},
\eeq

and 

\beq\label{fermi_vel}
\left(\frac{\del\veps_{p}}{\del p}\right)_{p=p_{f}}=v_{f}&=&
\frac{p_{f}}{\mu}-\frac{g_{s}g_{I}p_{f}^2}{2\pi^2}\frac{f_{1}}{3}.
\eeq

Now from Eq.(\ref{delmu_deln}) one gets

\beq\label{mudmu}
\mu{\rm d\mu}&=&\left[p_{f}+g_{s}g_{I}\frac{\mu p_{f}^2}{2\p^2}
(f_{0}-\frac{1}{3}f_{1})\right]dp_{f}.
\eeq

To calculate $\mu$, it is sufficient to let $\mu=\veps_{f}$ in the right hand 
side of Eq.(\ref{mudmu}). With the constant of integration adjusted so that
at high density $ p_{f}\simeq \veps_{f}$, Eq.(\ref{mudmu}) upon integration
together with Eq.(\ref{f0f1sig}) yield

\beq\label{mu_sigma}
\mu^{ex}_{\sigma}&=&\veps_{f}+\frac{g_{s}g_{I}g_{\sigma}^2}{128\pi^2\veps_{f}}M^{*2}
\left[-2y_{\sigma}(4-y_{\sigma}^2)^{3/2}\tan^{-1}\left
(\frac{x\sqrt{4-y_{\sigma}^2}}{y_{\sigma}\sqrt{1+x^2}}\right)\right.\nn\\&&\left.
~~~~~~~~~~~~~~~~~~~~~~~~~~~+\frac{y_{\sigma}^2(4-y_{\sigma}^2)\sqrt{1+x^2}}{x}
\ln\left(1+\frac{4x^2}
{y_{\sigma}^2}\right)+4x\sqrt{1+x^2}
\right.\nn\\&&\left.
~~~~~~~~~~~~~~~~~~~~~~~~~+2(y_{\sigma}^4-6y_{\sigma}^2+6)\ln(x+\sqrt{1+x^2})
\right],
\eeq

where $x=p_{f}/M^*$ and $y_{\s}=m_{\s}/M^*$.

\vskip 0.2in

\begin{figure}[htb]
\begin{center}
\resizebox{8cm}{6.0cm}{\includegraphics[]{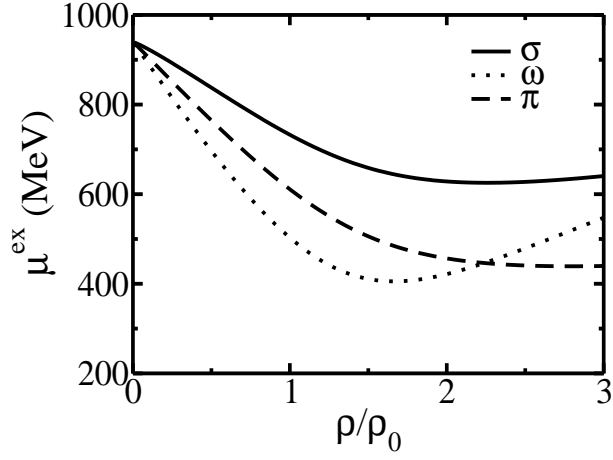}}
\caption{Density dependent of exchange chemical potential in symmetric nuclear
matter. $\s$,$\o$ and $\p$ mesons are denoted by solid, dotted and dashed line
respectively.}
\label{fig4}
\end{center}
\end{figure}

For massless meson limit {\em i.e.} at $m_{\s}\ra 0$ implies $y_{\s}\ra 0$ 
we have 

\beq
\mu^{ex}_{\s}\left\vert\right._{m_{\s}\ra 0}&=&\veps_{f}+g_{s}g_{I}
\frac{g_{\s}^2M^{*2}}{32\p^2\veps_{f}}\left[x\sqrt{1+x^2}+3\ln(x+\sqrt{1+x^2})
\right].
\eeq

Similarly for vector meson interaction, using Eq.(\ref{f0f1ome})
and Eq.(\ref{mudmu}) one obtains

\beq
\mu_{\omega}^{ex}&=&\veps_{f}-\frac{g_{s}g_{I}g_{\omega}^2M^{*2}}{64\pi^2\veps_{f}}
\left[
\frac{2y_{\omega}(y_{\omega}^4-2y_{\omega}^2-8)}{\sqrt{-y_{\omega}^2+4}}\tan^{-1}
\left(\frac{x\sqrt{-y_{\omega}^2+4}}{y_{\omega}\sqrt{1+x^2}}\right)\right.\nn\\
&&\left.~~~~~~~~~~~~~~~~~~~~~~~+\frac{y_{\omega}^2(y_{\omega}^2+2)\sqrt{1+x^2}}{x}
\ln\left(1+\frac{4x^2}{y_{\omega}^2}\right)-4x\sqrt{1+x^2}\right.\nn\\&&\left.
~~~~~~~~~~~~~~~~~~~~~~~~+2(6-y_{\omega}^4)\ln(x+\sqrt{1+x^2})\right].
\eeq

Here $y_{\o}=m_{\o}/M^*$. For massless limit of vector
meson the expression for chemical potential reads as

\beq
\mu^{ex}_{\o}&=&\veps_{f}+g_{s}g_{I}\frac{g_{\o}^2M^{*2}}{16\p^2\veps_{f}}
\left[x\sqrt{1+x^2}-3\ln(x+\sqrt{1+x^2})\right].
\eeq

In low density limit ($M^*\ra M$) for the massless meson exchange we reproduce
the expression derived earlier {\cite{baym76,chin77}}.

\vskip 0.2in
\begin{figure}[htb]
\begin{center}
\resizebox{8cm}{6.0cm}{\includegraphics[]{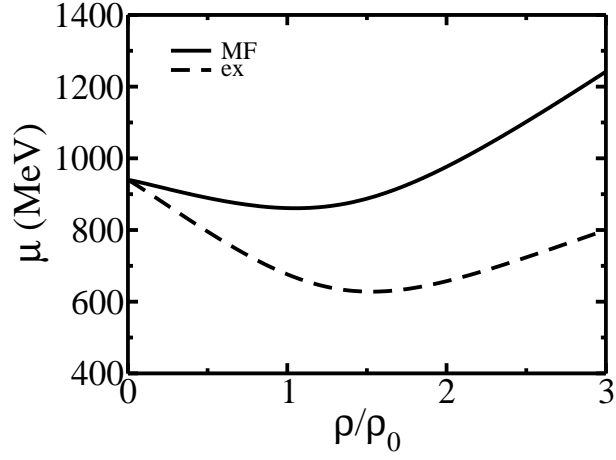}}
\caption{Comparison of mean field and exchange results of chemical potential 
in symmetric nuclear matter. Direct contributions are plotted by solid 
curve and exchange contributions by dashed curve.}
\label{fig5}
\end{center}
\end{figure}


\vskip 0.2in

\begin{figure}[htb]
\begin{center}
\resizebox{8cm}{6.0cm}{\includegraphics[]{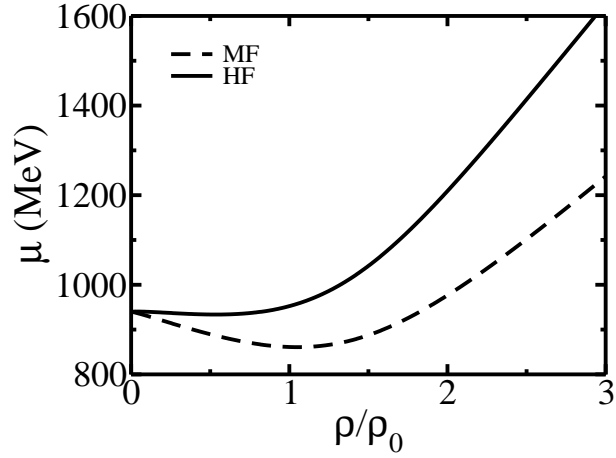}}
\caption{Comparison of chemical potential with MFT and HF case in symmetric 
nuclear matter. MF and HF results denoted by dashed and solid line 
respectively.}
\label{fig11}
\end{center}
\end{figure}


Once the $\mu^{ex}$ is determined, one can readily calculate its contribution 
to the energy density. For scalar meson interaction it is given by 
{\cite{baym76,chin77}},
 
\beq\label{xe_sig}
E^{ex}_{\s}&=&\frac{1}{2}\int{\rm d\rho}(\mu^{ex}_{\s}-\veps_{f})\nn\\
&=&g_{s}^2g_{I}^2\frac{g_{\s}^2M^{*4}}{512\p^4}
\left[y_{\s}^2(4-y_{\s}^2)\left\{-\frac{x^2}{2}+\frac{y_{\s}^2+4x^2}{8}
\ln\left(1+\f{4x^2}{y_{\s}^2}\right)\right\}+x^4\right.\nn\\
&&\left.~~~~~~~~~~~~~~~~~
-\f{1}{2}(y_{\s}^4-6y_{\s}^2+6)(\{x\sqrt{1+x^2}-\ln(x+\sqrt{1+x^2})\}^2-x^4)
+I_{\s}\right],
\eeq

where 

\beq
I_{\s}&=&-2y_{\sigma}(4-y_{\sigma}^2)^{3/2}\int\f{x^2}{\sqrt{1+x^2}}
\tan^{-1}\left(\frac{x\sqrt{4-y_{\sigma}^2}}{y_{\sigma}\sqrt{1+x^2}}\right)
{\rm d}x.
\eeq

Similarly, for vector meson exchange we obtain

\beq\label{xe_ome}
E^{ex}_{\o}&=&\frac{1}{2}\int{\rm d\rho}(\mu^{ex}_{\o}-\veps_{f})\nn\\
&=&-g_{s}^2g_{I}^2\frac{g_{\o}^2M^{*4}}{256\p^4}
\left[y_{\o}^2(2+y_{\o}^2)\left\{-\frac{x^2}{2}+\frac{y_{\o}^2+4x^2}{8}
\ln\left(1+\f{4x^2}{y_{\o}^2}\right)\right\}-x^4\right.\nn\\
&&\left.~~~~~~~~~~~~~~~~~
+\left(\f{y_{\o}^4}{2}-3\right)\left(\{x\sqrt{1+x^2}-\ln(x+\sqrt{1+x^2})\}^2-x^4
\right)+I_{\o}\right],
\eeq

where 

\beq
I_{\o}&=&\frac{2y_{\o}(y_{\o}^4-2y_{\o}^2-8)}{\sqrt{4-y_{\o}^2}}
\int\f{x^2}{\sqrt{1+x^2}}
\tan^{-1}\left(\frac{x\sqrt{4-y_{\o}^2}}{y_{\o}\sqrt{1+x^2}}\right)
{\rm d}x.
\eeq

Thus for the case of massless mesons {\cite{baym76,chin77}}, we have

\beq\label{sig_eng_mless}
E^{ex}_{\s}|_{m_{\s}\ra 0}&=&\f{g_{\s}^2M^{*4}}{8\p^4}
[x^4-\f{3}{4}\{x\sqrt{1+x^2}-\ln(x+\sqrt{1+x^2})\}^2],
\eeq

and

\beq\label{omg_eng_mless}
E^{ex}_{\o}|_{m_{\o}\ra 0}&=&-\f{g_{\o}^2M^{*4}}{8\p^4}
[x^4-\f{3}{2}\{x\sqrt{1+x^2}-\ln(x+\sqrt{1+x^2})\}^2].
\eeq

Due to presence of pion fields in the chiral Lagrangian we have component in 
the interaction which acts on the isospin fluctuation. One can derive the 
quasiparticle interaction with isospin dependency by the same procedure as for 
$\sigma$ and $\omega$ meson. Pion, being a pseudoscalar, fails to contribute 
at the MF level forcing us to go beyond MFT so as to include pionic 
contribution to the FLPs. It is to be noted that, in exchange diagram 
pion have both isoscalar and isovector contribution to FLPs. Detailed 
calculation for isoscalar contribution to FLPs is similar as $\sigma$ and 
$\omega$ meson. For brevity, we present only dimensionless FLPs and their contribution to energy density. We also quote their numerical values.

The dimensionless LPs due to $\pi$ exchange are

\beq\label{dless_F0_pi}
F^{ex,\p}_{0}=-g_{s}g_{I}\f{3g_{A}^2p_{f}M^{*2}}{64\p^2f_{\p}^2\veps_{f}}
\lt[-2+\frac{m_{\pi}^2}{2p_{f}^2}
\ln\left(1+\frac{4p_{f}^2}{m_{\pi}^2}\right)\rt],
\eeq

and

\beq\label{dless_F1_pi}
\f{1}{3}F^{ex,\p}_{1}=-g_{s}g_{I}\f{3g_{A}^2m_{\p}^2M^{*2}}
{128\p^2f_{\p}^2p_{f}\veps_{f}}
\lt[-2+\left(\frac{m_{\pi}^2}
{2p_{f}^2}+1\right)\ln\left(1+\frac{4p_{f}^2}{m_{\pi}^2}\right)\rt].
\eeq


\begin{figure}[tb]
\begin{center}
\resizebox{8cm}{6.0cm}{\includegraphics[]{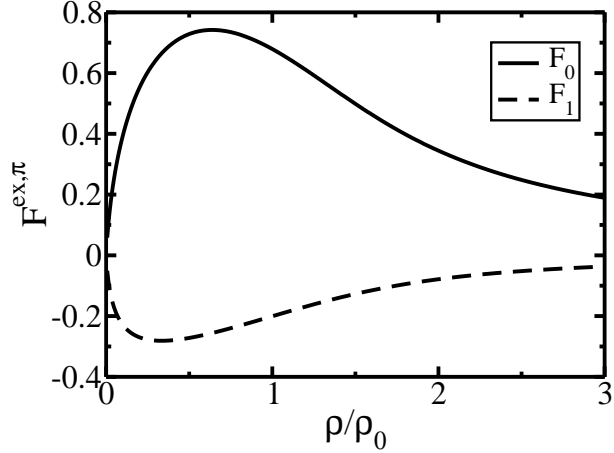}}
\caption{Dimensionless LPs in symmetric nuclear 
matter for pion exchange in relativistic theory. 
Solid and dashed line represent $F_{0}^{\pi}$ and $F_{1}^{\pi}$ respectively.}
\label{fig8}
\end{center}
\end{figure}



\begin{figure}[tb]
\begin{center}
\resizebox{8cm}{6.0cm}{\includegraphics[]{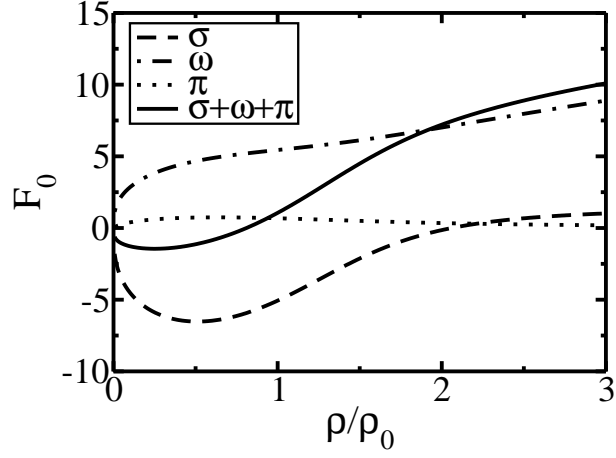}}
\caption{Dimensionless relativistic LP $F_{0}$ in symmetric 
nuclear matter. $\s$, $\o$, $\p$ and total contribution are denoted by
dashed, dot-dashed, dotted and solid line respectively.}
\label{fig9}
\end{center}
\end{figure}



\begin{figure}[tb]
\begin{center}
\resizebox{8cm}{6.0cm}{\includegraphics[]{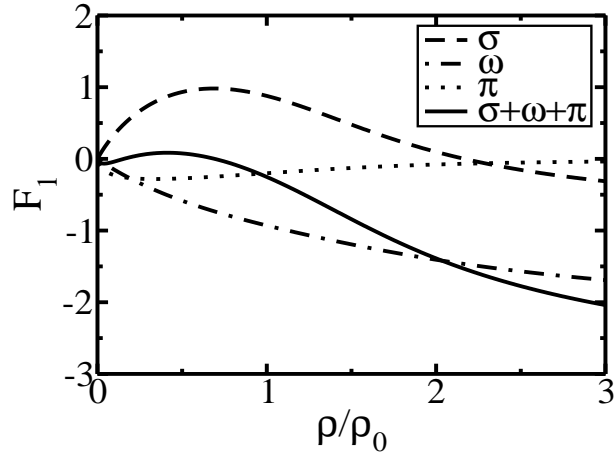}}
\caption{Dimensionless relativistic LP $F_{1}$ in symmetric 
nuclear matter. $\s$, $\o$, $\p$ and total contribution are denoted by
dashed, dot-dashed, dotted and solid line respectively.}
\label{fig10}
\end{center}
\end{figure}


In the non-relativistic limit $\veps_{f}\ra M^*$, one obtains the same
expression of $F_{1}^{\pi}$ as reported in \cite{friman96,rho80}.
In Fig.(\ref{fig8}) we show the density dependence of $F_{0}$ and 
$F_{1}$ due to pionic interaction. Numerically 
at nuclear saturation density $(\rho_{0}=0.148 {\rm fm^{-3}})$, 
$F^{ex,\p}_{0}=0.68$ and $F^{ex,\p}_{1}=-0.2$.

In Fig.(\ref{fig9}) and Fig.(\ref{fig10}) we plot separate and total 
contribution of $F_0$ and $F_1$ due to $\s$, $\o$ and $\p$ exchange respectively.
Interestingly, individual contribution to LPs of $\s$ and $\o$ meson are
large while sum of their contribution to $F_{0}^{tot}$ is small due to the
sensitive cancellation of $F_{0}^{\s}$ and $F_{0}^{\o}$ as can be seen
from Fig.(\ref{fig9}). Such a cancellation is responsible for the nuclear saturation dynamics  \cite{celenza_book,anastasio83}.  
Numerically, $F_{0}^{\s+\o}$ is approximately $3/2$ 
times smaller than $F_{0}^\p$ as can be seen from Table (\ref{table-3}).


\begin{figure}[!tb]
\begin{center}
\resizebox{8cm}{6.0cm}{\includegraphics[]{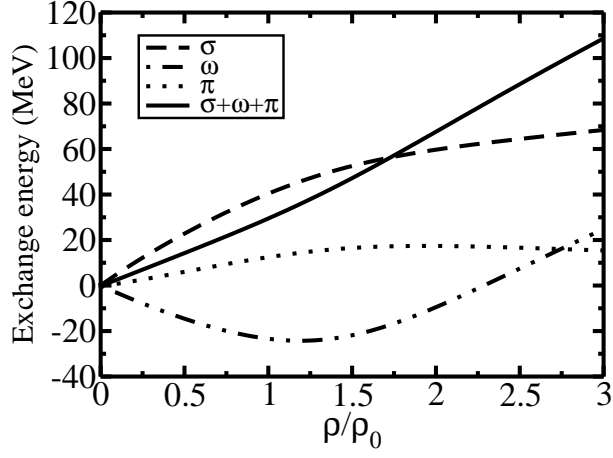}}
\caption{Comparison of separate and total exchange energy obtained from FLT in 
symmetric nuclear matter.$\s$, $\o$, $\p$ and total contribution are denoted by
dashed, dot-dashed, dotted and solid line respectively.}
\label{fig12}
\end{center}
\end{figure}



\begin{figure}[tb]
\begin{center}
\resizebox{8cm}{6.0cm}{\includegraphics[]{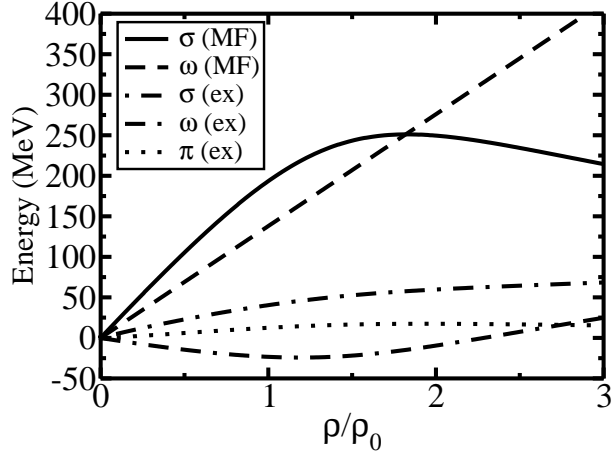}}
\caption{Comparison of mean field energy and exchange energy obtained from FLT 
in symmetric nuclear matter.
$E_{\s}^{MF}$, $E_{\o}^{MF}$, $E_{\s}^{ex}$, $E_{\o}^{ex}$ and $E_{\p}^{ex}$ are 
denoted by solid, 
dashed, dot-dashed, dash-dashed and dotted line respectively.}
\label{fig13}
\end{center}
\end{figure}


The exchange energy density is given by 

\beq\label{xe_pi}
E^{ex}_{\p}&=&-g_{s}^2\frac{3g_{A}^2M^{*6}}{128f_{\p}^2\p^4}
\left[I_{\p}+y_{\p}^4\left\{-\frac{x^2}{2}+\frac{y_{\p}^2+4x^2}{8}
\ln\left(1+\f{4x^2}{y_{\p}^2}\right)\right\}-x^4\right.\nn\\
&&\left.
+\left(\f{y_{\p}^4}{2}-y_{\p}^2-1\right)
\left(\{x\eta-\ln(x+\eta)\}^2-x^4\right)\right],
\eeq

where $y_{\p}=m_{\p}/M^*$ and 

\beq
I_{\p}=-2y_{\p}^3\sqrt{4-y_{\p}^2}
\int\f{x^2}{\eta}
\tan^{-1}\left(\frac{x\sqrt{4-y_{\p}^2}}{y_{\p}\eta}\right)
{\rm d}x.
\eeq

For the massless pion this reads as

\beq\label{pi_eng_mless}
E^{ex}_{\p}{\Big\vert}_{m_{\p}\ra 0}&=&\f{3g_{A}^2M^{*6}}{32f_{\p}^2\p^4}
[x\eta-\ln(x+\eta)]^2.
\eeq

In Fig.(\ref{fig12}) and Fig.(\ref{fig13}) we show the density dependence 
of energy due $\s$, $\o$ and $\p$ meson exchanges. Numerical values are
quoted in Table(\ref{table-5}).

It might be mentioned here that in the massless meson limit, 
Eq.(\ref{sig_eng_mless}), (\ref{omg_eng_mless}) and (\ref{pi_eng_mless})
can be evaluated analytically from two loop ring diagrams of ref.\cite{hu07} 
using Eqs.(54), (55) and (56). We have checked and the expression for 
the energies are found to be consistent with each other.
With massive meson results are compared numerically. 
Numerical estimation of exchange energy 
from loop diagram and RFLT are found to be few percent limit.


\begin{table}
\caption{Dimensionless Landau parameters and chemical potential at 
$\rho=\rho_{0}$. Note that, $F_{0}$, $F_{1}$ are the dimensionless 
isoscalar LPs.}
\label{table-3}
\begin{center}
\begin{tabular}{ccccccc} \hline \hline
 Meson   & $F_{0}$ & ~~~~~~ & $F_{1}$  &  ~~~~~~  & $\mu^{ex}(MeV)$\\ \hline
  $\s$   &  -5.04  & ~~~~~~ & 0.875    &  ~~~~~~  & 731.89 \\
  $\o$   &   5.44  & ~~~~~~ & -0.93    &  ~~~~~~  & 501.82 \\
  $\p$   &   0.68   & ~~~~~~ & -0.20   &  ~~~~~~  & 609.88  \\ \hline\hline
\end{tabular}
\end{center}
\end{table}



\begin{table}
\caption{Chemical potential in MeV from FLT at $\rho=\rho_{0}$.}
\label{table-4}
\begin{tabular}{ccccccc} \hline \hline
 Meson    &~~~~   & $\mu^{ex}$ &~~~~  & $\mu^{MF}$ &~~~~ &$\mu^{HF}$\\ \hline
  $\s$    &~~~~   &  731.89   &~~~~   &    -      &~~~~ &    -     \\
  $\o$    &~~~~   &  501.82   &~~~~   &    -      &~~~~ &    -      \\ 
  $\p$    &~~~~   &  609.88   &~~~~   &    -      &~~~~ &    -      \\     
  $\s+\o+\p$ &~~~~ & 675.63   &~~~~   &  861.07   &~~~~ &  952.73    \\  
\hline\hline
\end{tabular}
\end{table}


Finally we reproduce saturation property of nuclear matter {\em i.e.}
$E/{\rho}-M=-16.12$ MeV at $p_{f}=1.3 {\rm fm^{-1}}$ with those energy 
calculated from RFLPs.



\begin{table}
\caption{MF and Exchange energy in MeV from FLT at $\rho=\rho_{0}$.}
\label{table-5}
\begin{tabular}{ccccc} \hline \hline
 Meson    &~~~~    & $E^{MF}$      &~~~~    & $E^{ex}$    \\  \hline
  $\s$    &~~~~    &  193.86      &~~~~    &  40.48      \\
  $\o$    &~~~~    &  138.39      &~~~~    & -23.41       \\ 
  $\p$    &~~~~    &    -         &~~~~    &  12.49       \\    \hline\hline
\end{tabular}
\end{table}



\begin{figure}[ht]
\vskip 0.15in
\begin{center}
\resizebox{8.0cm}{6.0cm}{\includegraphics[]{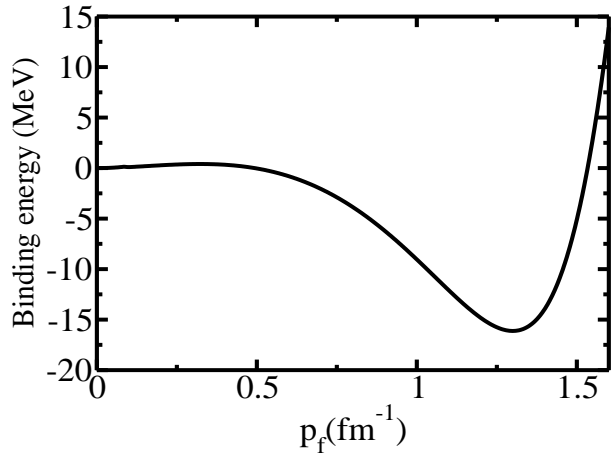}}
\caption{Binding energy graph from FLT for symmetric nuclear matter.}
\label{fig15}
\end{center}
\end{figure}



\vskip 0.4in
\subsection{Incompressibility and First Sound Velocity}

In nuclear matter several important relationships exist between nuclear 
observables and the FLPs. The thermodynamical parameters
can be expressed in terms of few LPs.
For example we present the incompressibility ($K$) and 
first sound velocity ($c_{1}$)\cite{holt07,matsui81}.

\vskip 0.2in
\begin{figure}[htb]
\begin{center}
\resizebox{8cm}{6.0cm}{\includegraphics[]{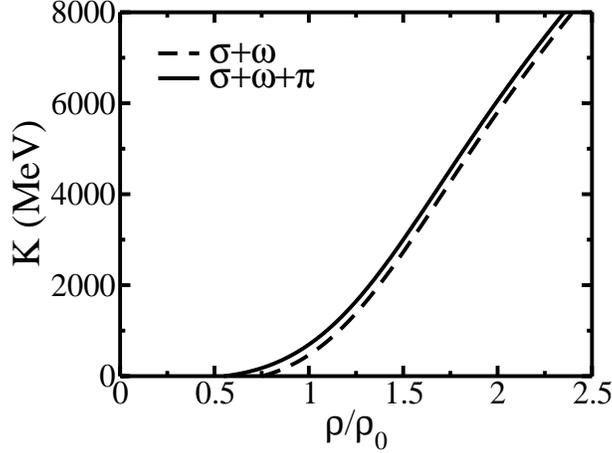}}
\caption{Compressibility $K$ in symmetric nuclear matter. 
}
\label{fig6}
\end{center}
\end{figure}


Incompressibility of the Fermi liquid may be derived as in the 
non-relativistic theory by the second derivative of energy density ($E$) with 
respect to the number density $\rho$ {\cite{holt07,matsui81}};

\beq
K&\equiv&\rho\frac{\del^2 E}{\del \rho^2}\nn\\
&=&\rho\frac{\del\mu}{\del \rho}.
\eeq

If energy density $E$ is given in terms of number density $\rho$, then the 
expression for incompressibility or compression modulus in terms of 
LPs is given by,

\beq\label{incompressibility}
K&=&\frac{3p_{f}^2}{\veps_{f}}(1+F_{0}).
\eeq

Now consider the effect of quasiparticle collision on the collective modes of
a neutral Fermi liquid. Suppose the frequency of the mode is $\omega$, while 
the quasiparticle collision frequency is $\nu$. For the limit $\omega<<\nu$, 
many quasiparticle collision takes place during time interval $\omega^{-1}$.
Then the region is collision-dominated, or {\it  hydrodynamic regime}
{\cite{matsui81}}. Under this circumstances, organized density 
fluctuation is possible and  hydrodynamic or first sound waves are 
generated. Hydrodynamic sound propagates only when the
system attains the local thermodynamic equilibrium in a time much shorter
than the time interval of the sound oscillation.


\begin{figure}[tb]
\begin{center}
\resizebox{8cm}{6.0cm}{\includegraphics[]{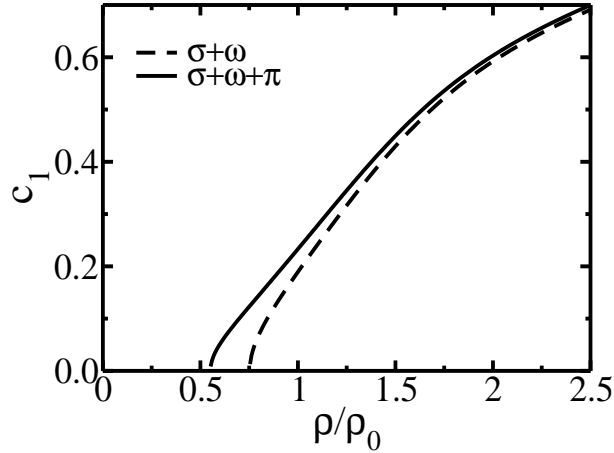}}
\caption{First sound velocity $c_{1}$ in symmetric nuclear matter. 
}
\label{fig7}
\end{center}
\end{figure}


The first sound velocity is given by {\cite{baym76}}

\beq\label{first_sound}
c_{1}^2~=~\frac{\del P}{\del E}&=&\frac{\del P}{\del \mu}\frac{\del \mu}
{\del \rho}\frac{\del \rho}{\del E}\nn\\
&=&\frac{\rho}{\mu}\frac{\del \mu}{\del \rho},
\eeq

With the help of Eq.(\ref{delmu_deln}), Eq.(\ref{first_sound}) yields

\beq\label{first_sound_1}
c_{1}^2&=&\frac{1}{3}\frac{p_{f}^2}{\mu^2}\frac{1+F_{0}}{1+\frac{1}{3}F_{1}}\nn\\
&=&\frac{1}{3}\frac{p_{f}^2}{\mu^2}\left[1+\frac{g_{s}g_{I}\mu p_{f}}{2\pi^2}
(f_{0}-\frac{1}{3}f_{1})\right].
\eeq

Corresponding values of the incompressibility and the first
sound velocity are plotted in Fig.(\ref{fig6}) and Fig.(\ref{fig7}) 
separately with $\s+\o$ and $\s+\o+\p$ contribution. It is 
observed that for combined $\s$ and $\o$ meson at $\rho<0.75\rho_{0}$, 
$F_{0}<-1$ and the resulting compressibility turns out to be negative. 
While for $\pi$ meson together with $\s$ and $\o$ meson the same conclusion 
can be drawn at $\rho<0.55\rho_{0}$ . This is the region where 
the attractive interaction due to the exchange of scalar mesons overwhelms 
the repulsive force coming from vector meson exchange, and consequently the 
system becomes unstable {\cite{matsui81}}. On the other hand, as the density 
increases, the attractive scalar meson exchange force tends to be suppressed 
by the relativistic effect and the net quasiparticle interaction become 
repulsive. At nuclear saturation density ($\rho_{0}=0.148 {\rm fm^{-3}}$) 
we have $K=476.04$ MeV and $705.84$ MeV for combined $\s+\o$ and 
$\s+\o+\p$ mesons respectively. The small effective mass is responsible 
for large incompressibility. The first sound velocity $c_{1}=0.19$ 
for $\s+\o$ and $0.23$ for $\s+\o+\p$ at normal nuclear matter 
density and at all densities $c_{1}$ is smaller than the velocity of 
light, which is consistent with causality {\cite{matsui81}}.


\vskip 0.4in
\section {Isovector Landau parameter and symmetry energy}

In this section we proceed to calculate isovector LPs due to pion exchange. 
The isovector contribution to interaction parameter is

\beq\label{pion_interaction}
f_{pp'}^{\prime}{\Big\vert}_{p\simeq p'=p_{f}}&=&
-\frac{1}{2}\f{g_{A}^2M^{*2}}{4f_{\p}^2\veps_{f}^2}
\lt\{\f{p_{f}^2(1-\cos\th)}{2p_{f}^2(1-\cos\th)+m_{\p}^2}\rt\}.
\eeq

where $g_{A}^2=1.5876$ , $f_{\p}=93 MeV$  and $m_{\p}=139$ MeV {\cite{hu07}}. 
Here $-1/2$ is isospin factor in isovector channel {\cite{friman99}}.

Using Eq.(\ref{landau_para}) and (\ref{pion_interaction}) we derive 
isovector LPs $f'_{0}$ and $f'_{1}$,

\beq\label{f0_pi}
f^{\prime}_{0}&=&-\f{g_{A}^2M^{*2}p_{f}^2}{16f_{\p}^2\veps_{f}^2}
\int_{-1}^{1}\f{(1-\cos\theta)}{2p_{f}^2(1-\cos\theta)+m_{\p}^2}
{\rm d(\cos\theta)}\nn\\
&=&\f{g_{A}^2M^{*2}}{32f_{\p}^2\veps_{f}^2}
\left[-2+\frac{m_{\pi}^2}{2p_{f}^2}
\ln\left(1+\frac{4p_{f}^2}{m_{\pi}^2}\right)\right],
\eeq

and

\beq\label{f1_pi}
\frac{1}{3}f^{\prime}_{1}&=&-\f{g_{A}^2M^{*2}p_{f}^2}{16f_{\p}^2\veps_{f}^2}
\int_{-1}^{1}\f{\cos\theta(1-\cos\theta)}{2p_{f}^2(1-\cos\theta)+m_{\p}^2}
{\rm d(\cos\theta)}\nn\\
&=&\f{g_{A}^2M^{*2}m_{\p}^2}
{64f_{\p}^2\veps_{f}^2p_{f}^2}
\left[-2+\left(\frac{m_{\pi}^2}
{2p_{f}^2}+1\right)\ln\left(1+\frac{4p_{f}^2}{m_{\pi}^2}\right)\right].
\eeq

The dimensionless LPs $F_{0}^{\prime}=N(0)f_{0}^{\prime}$ and 
$F_{1}^{\prime}=N(0)f_{1}^{\prime}$ are given below. Here, 
$N(0)$ is the density of states at the Fermi surface defined in 
Eq.(\ref{dens_of_state}).  
 
\beq
F^{\prime}_{0}&=&g_{s}g_{I}\f{g_{A}^2p_{f}M^{*2}}{64\p^2f_{\p}^2\veps_{f}}
\lt[-2+\frac{m_{\pi}^2}{2p_{f}^2}
\ln\left(1+\frac{4p_{f}^2}{m_{\pi}^2}\right)\rt],
\eeq

and

\beq
\f{1}{3}F^{\prime}_{1}&=&g_{s}g_{I}\f{g_{A}^2m_{\p}^2M^{*2}}
{128\p^2f_{\p}^2p_{f}\veps_{f}}
\lt[-2+\left(\frac{m_{\pi}^2}
{2p_{f}^2}+1\right)\ln\left(1+\frac{4p_{f}^2}{m_{\pi}^2}\right)\rt].
\eeq


\begin{figure}[tb]
\begin{center}
\resizebox{8cm}{6.0cm}{\includegraphics[]{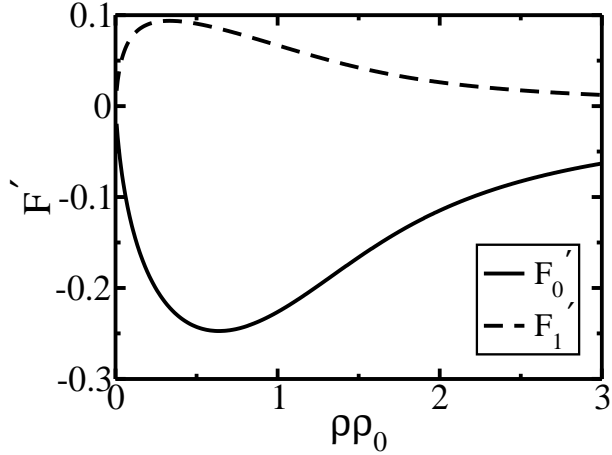}}
\caption{Density dependent of isovector LPs 
in symmetric nuclear matter.}
\label{fig14}
\end{center}
\end{figure}


Knowing the isovector LPs, to which here only the pion contributes,
one can calculate nuclear symmetry energy. The symmetry
energy is defined as the difference of energy between the neutron matter and
symmetric nuclear matter which we denote as $\beta$ \cite{matsui81}. 

Analytically, the symmetry energy is defined 
from the expansion of the energy per nucleon $E'(\rho,\alpha)$ in terms of the
asymmetry parameter $\alpha$ defined as {\cite{greco03}}

\beq
\alpha \equiv-\frac{\rho_{3}}{\rho}=\frac{\rho_{n}-\rho_{p}}{\rho}=\frac{N-Z}{A}.
\eeq

We have

\beq
E'(\rho,\alpha)\equiv\frac{E(\rho,\alpha)}{\rho}=E'(\rho)+E'_{sym}(\rho)\alpha^2
+{\cal{O}}(\alpha^4)+.....
\eeq

and so, in general,

\beq
E'_{sym}\equiv\beta&=&\frac{1}{2}\frac{\del^{2}E'(\rho,\alpha)}{\del\alpha^2}
{\Big|}_{\alpha=0}\nn\\
&=&\frac{1}{2}\rho\frac{\del^{2}E}{\del\rho_{3}^2}{\Big|}_{\rho_{3}=0}.
\eeq

In terms of LPs, the symmetry energy can be expressed as 

\beq\label{symm_energy}
\beta = \frac{p_{f}^2}{6\veps_{f}}(1+F'_{0}),
\eeq

where $F'_{0}$ is the isospin dependent LP. Numerically at saturation 
density $(\rho=\rho_0)$ we obtain $\beta=14.57$ MeV. So relatively small contribution to $\beta$ comes from one-pion exchange diagram {\cite{die03}}.

Similar to the hydrodynamic sound corresponding to the total baryon density oscillation, we consider hydrodynamic isospin sound which accompany the 
out-of-phase oscillations of proton and neutron density. Isospin sound 
velocity $c_{1}^{\prime}$ is given by {\cite{matsui81}}

\beq\label{iso_sound}
c_{1}^{\prime}&=& v_{f}\sqrt{\frac{\veps_{f}}{3\mu}(1+F_{0}^{\prime})}
\eeq

Numerically at saturation density
$(\rho=\rho_0)$ we obtain $c_1^{\prime}=0.17$.


\vskip 0.4in
\section{Discussion}

In this work we determine the RFLPs using effective chiral model to study 
the properties of DNM. We compare the MF results with the corresponding 
perturbative results for the FLPs. It is seen at low densities, 
they converge but perturbative results differs significantly with the
mean field results at higher density.

We have estimated the pionic contribution to the FLPs which can contribute
only in exchange diagrams. It is seen that numerically pionic contribution is 
smaller than the corresponding $\sigma$ and $\omega$ meson results, however,
we show this is still important in various physical quantities as $\sigma$
and $\omega$ meson contribute in opposite sign and pionic contribution is
comparable to the sum of the contributions coming from the scalar and
vector meson exchanges. Among the other physical quantities
we also evaluate the sound velocity and incompressibility and symmetry 
energy etc. The results for the sound velocities gives the correct causal
limit at extremely high densities. It is also shown that, the system 
is stable with respect to density fluctuation at and above the nuclear 
saturation density and exhibit instability in relatively low density
regimes.

Here we also have calculated exchange energy densities by using FLPs 
calculated relativistically. The results are found to be consistent 
with what are obtains by directly evaluating the nucleon loops 
\cite{hu07}, this gives further support to the present approach. 

\vskip 0.2in
{\bf Acknowledgments}\\

The authors would like to thank G.Baym and C.Gale for their valuable 
comments. We also wish to thank P.Roy for his 
critical reading of the manuscript.



\end{document}